\input psfig
\documentstyle[12pt,epsf]{article}
\begin{document}
\setlength{\baselineskip}{0.33 in}
\catcode`@=11
\long\def\@caption#1[#2]#3{\par\addcontentsline{\csname
  ext@#1\endcsname}{#1}{\protect\numberline{\csname
  the#1\endcsname}{\ignorespaces #2}}\begingroup
    \small
    \@parboxrestore
    \@makecaption{\csname fnum@#1\endcsname}{\ignorespaces #3}\par
  \endgroup}
\catcode`@=12
\newcommand{\newc}{\newcommand}
\newc{\gsim}{\lower.7ex\hbox{$\;\stackrel{\textstyle>}{\sim}\;$}}
\newc{\lsim}{\lower.7ex\hbox{$\;\stackrel{\textstyle<}{\sim}\;$}}
\newenvironment{fnitemize}{\begin{itemize}
	\footnotesize}{\end{itemize}}

\vspace*{-1.25in}
\vspace*{.85in}
\begin{center}
{\Large{\bf  Loop Representations of the Quark Determinant in Lattice QCD}
}\\
\vspace*{.45in}
{\large{A.~Duncan$^1$, E.~Eichten$^2$,  R.~Roskies$^1$, and H.~Thacker$^3$}} \\ 
\vspace*{.15in}
$^1$Dept. of Physics and Astronomy, University of Pittsburgh, Pittsburgh, PA 15620\\
$^2$Fermilab, P.O. Box 500, Batavia, IL 60510 \\
$^3$Dept.of Physics, University of Virginia, Charlottesville, VA 22901
\end{center}
\vspace*{.3in}
\begin{abstract}
 The modelling of the ultraviolet contributions to the quark determinant in
 lattice QCD in terms of a small number of Wilson loops is examined. Complete 
 Dirac spectra are obtained for sizeable ensembles of SU(3) gauge fields at $\beta = 5.7$
 on 6$^4$, 8$^4$ and 10$^4$ lattices allowing for the first time a detailed study of the  volume
 dependence of the effective loop action generating the quark determinant.
 The connection to the hopping parameter expansion is examined in the
 heavy quark limit. We compare the efficiency and accuracy of various methods - specifically, Lanczos
 versus stochastic approaches- for extracting  the quark determinant
 on an ensemble of configurations.
\end{abstract}

\newpage

\section{Introduction}

In a recent paper \cite{fullqcd} we introduced a method for
performing an unquenched Monte Carlo
simulation in lattice QCD in which the infrared and ultraviolet modes
of the quark fields are
treated separately. The low eigenvalues (typically up to a cutoff
somewhat above $\Lambda_{QCD}$) are exactly and explicitly
calculated and included as a truncated quark determinant in the the update
Boltzmann measure
The remaining UV modes are included approximately by
modelling the high end of the spectrum with an effective loop action involving only
small Wilson loops. It was found that on small lattices (specifically, $6^4$ lattices
at $\beta$=5.7) the accuracy of such a loop fit to the high end of the
determinant was remarkably good- the $\chi^2$
per degree of freedom of  a fit with loops of up to 6 links to the logarithm
of the quark determinant, including all
modes above 340 MeV was  0.23, while the typical excursion of the log determinant  between
uncorrelated configurations was on the order of 20.  
Of course, one expects that the
accuracy of such a fit will decrease with increasing volume,  and it is not clear 
that this approach
will remain practical  once lattices of physically useful size are reached.  As an
illustration, we note
that recent studies of electromagnetic effects in lattice QCD 
\cite{empapers} found that the finite volume effects from even the long-range electromagnetic
effects were controllable  on 12$^3$x24 lattices at $\beta$=5.9.  A 10$^3$x20 lattice
at $\beta$=5.7 has almost twice the physical volume, while the lattice discretization
effects can presumably be substantially reduced by using clover improvement. Our aim in
this paper is therefore to study the accuracy of effective loop action representations
to infrared truncated quark determinants for various size lattices at $\beta$=5.7.  We
shall show that reasonably accurate
representations of the UV contribution to the quark determinant in terms
of a small number of Wilson loops are indeed possible on such physically useful lattices.
To the extent that a small residual error in the loop representation of the ultraviolet 
part of the quark determinant contributes primarily to an overall rescaling of dimensional 
quantities such as ground-state hadron masses (which
are dominated by quarks with a limited range of offshellness)
such a representation should be perfectly adequate for
dynamical spectrum calculations on moderately sized lattices.
 
  Past studies of loop representations of the quark determinant \cite{SextWein,IrvSext}
(which typically focussed on the {\em complete} determinant, less accurately
described by small loops) have been hampered by the difficulty of obtaining exact
values  for the quark determinant for a sufficiently large sample of independent gauge
configurations. Stochastic methods \cite{golub1, golub2} can be applied to fairly large
lattices but with limited accuracy, whereas the direct Lanczos approach \cite{lanczos}
loses steam for lattices larger than about 12$^4$. In this paper we have employed
the Lanczos approach to obtain complete exact Wilson-Dirac spectra for sizeable
ensembles of 6$^4$, 8$^4$ and 10$^4$ lattices, allowing us to study in detail the volume
dependence (Section 2) and quark mass dependence (Section 3) of the effective
loop action fit to the truncated quark determinant at various infrared cutoffs. Technical
details of the calculations, as well as a comparison of the computational burden of
the exact Lanczos and stochastic approaches, are presented in Section 4.
 
\section{Loop Actions for light quarks- volume dependence}

  The main difficulty we encounter in determining the accuracy of  a loop representation
for full or truncated quark determinants in lattice QCD lies in the computational
effort required to extract complete spectra of the Wilson-Dirac operator for a
sufficiently large sample of independent gauge configurations on lattices large
enough to yield physically useful information. A variety of  numerical tools, both
exact and statistical, now exist for accomplishing this task. Technical details of the
implementaion of these methods will be deferred to Section 4. In this section we
present a detailed study of the volume and IR-cutoff dependence of loop fits to the quark   
determinant for ensembles of  75 6$^4$  lattices, 75 8$^4$  lattices and 30 10$^4$  lattices
at $\beta=$5.7 and $\kappa=$0.1685. For all of these configurations we have carried out 
a complete spectral resolution using the Lanczos methods discussed in Section 4, 
and identifying for each configuration all
15552 (resp. 49152, 120000) eigenvalues in the 6$^4$ (resp. 8$^4$, 10$^4$) cases.
Most of these calculations were performed on a 9 node Beowulf system running Linux. 

  The configurations used in this study were generated using the truncated determinant
algorithm of  \cite{fullqcd}. In particular, the update measure included the contributions
to the quark determinant from low eigenmodes of  the hermitian Wilson-Dirac
operator $\gamma_{5}(D_W - m)$ up to a cutoff $\Lambda_{QCD}$ (specifically, we chose
a cutoff of 0.45 in lattice units, corresponding to about 490 MeV in physical
units). This cutoff corresponds to the lowest  30 (15 positive and 15 negative)
eigenvalues for the 6$^4$ lattices, and to 120 (resp. 350) low eigenvalues for the 8$^4$
(resp. 10$^4$) lattices. 
It is reasonable to expect that, by including this infrared contribution in the 
simulation measure the low energy chiral
structure is properly treated \cite{fullqcd}. The essence of the task being addressed
in this paper is then to determine the extent to which the remaining omitted ultraviolet
modes can accurately be fit by a gauge-invariant loop expansion involving relatively
few and simple loops. The loop actions discussed here will include only loops (or
Polyakov lines)  with up to 6 links, although the inclusion of loops with 8 links is completely
straightforward in principle and would of course give a substantial increase in the
accuracy of the fit. 
\begin{figure}
\psfig{figure=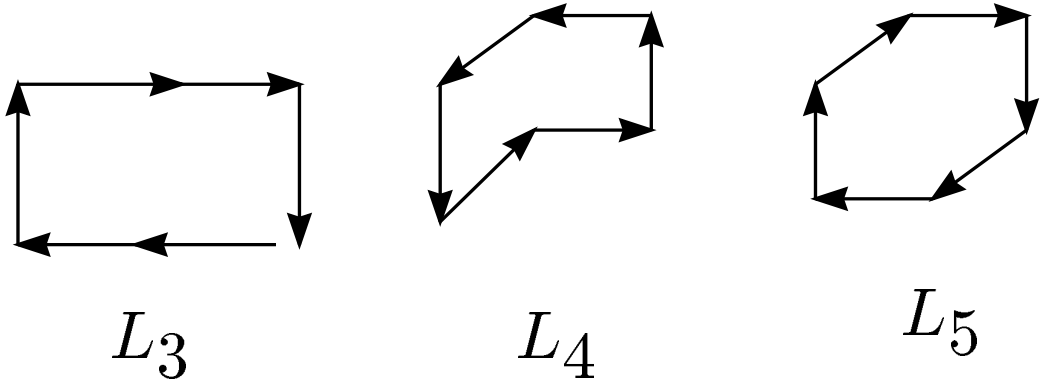,
width=0.95\hsize}
\caption{6 link closed loop contributions to $\cal{D}$}
\label{fig:loops}
\end{figure}

  On a 6$^4$ lattice with
periodic boundary conditions, there are five independent gauge invariant contributions
to $\cal{D}$=$\ln{{\rm det}(\gamma_{5}(D_W - m))}$ involving 6 links or less, 
corresponding to the plaquette (1x1 loop), here denoted $L_{1}(U)$,
the length 6 Polyakov line traversing the full length of the lattice in any direction, $L_{2}(U)$,
and 3 independent length 6 closed curves (denoted $L_{3,4,5}(U)$), illustrated
in Fig.1. In the case of the 8$^4$ and 10$^4$ lattices there are only  four such
terms- the plaquette
and the length 6 simply connected loops displayed in Fig. 1. In addition the fits contain
a constant term which can be regarded as the lattice equivalent of the unit operator.
We shall be studying the ultraviolet contribution to $\cal{D}$ in which the $n_{\lambda}$
lowest modes are omitted:
\begin{equation}
  {\cal D}(n_{\lambda})\equiv \sum_{i>n_{\lambda}} \ln{(|\lambda_{i}|)}
\end{equation}
where $\lambda_{i}$ are the gauge-invariant eigenvalues of  $(\gamma_{5}(D_W - m))$
enumerated in order of increasing absolute value.
Denoting the approximate loop value of ${\cal D}$ by $S_a$ and the true value by $S_t$, 
the variance per degree of freedom of the fit, $\sigma^2$, is defined as:
\begin{equation}
  \sigma^2 = \sum _{n=1}^{n=nc} (S_a(n)-S_t(n))^2  /(nc-np).
\end{equation}
for $nc$ configurations and $np$ loop variables (including the constant term).
(We note here
that the terminology "variance per degree of freedom" replaces the usual "chi-squared per
degree of freedom" as we are dealing with a dimensionless quantity 
without statistical errors.)
In Figures (2-4) we show the accuracy of the best fit to ${\cal D}(n_{\lambda})$, for
the cutoff corresponding to the actual simulation values - namely $n_{\lambda}=$30
(resp. 120, 350 for the 6$^4$ (resp. 8$^4$, 10$^4$) lattices. 
The $\sigma^{2}$ is 0.24, 0.28 and 0.79 for the 3 cases studied .
The very close matching of the
loop action to the determinant values for $\sigma^{2} <$1 suggests that accurate dynamical 
calculations should be possible by replacing the UV part of the quark determinant by such
a loop Ansatz.

\begin{figure}
\psfig{figure=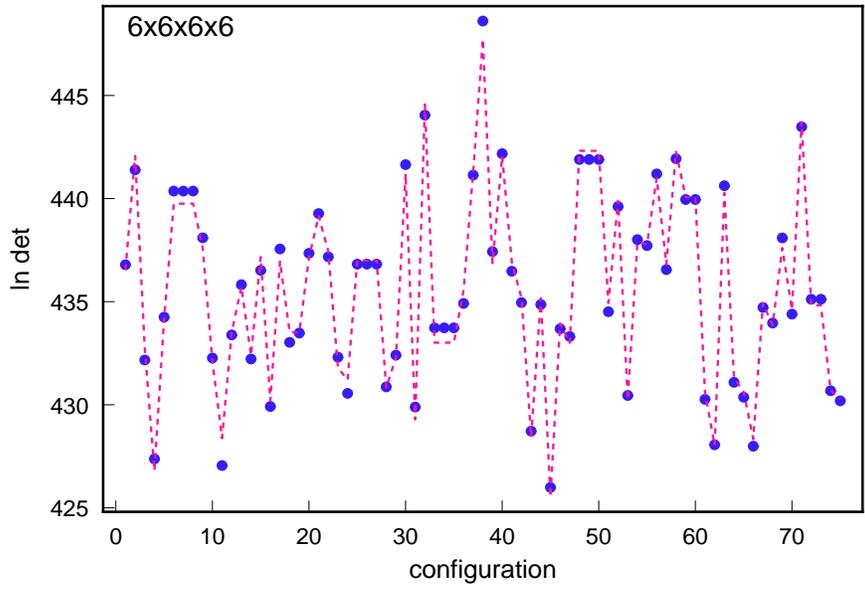,
width=0.8\hsize}
\caption{6 link fit  to ${\cal D}_{30}$, 6$^4$ lattices}
\label{fig:loops}
\end{figure}
\begin{figure}
\psfig{figure=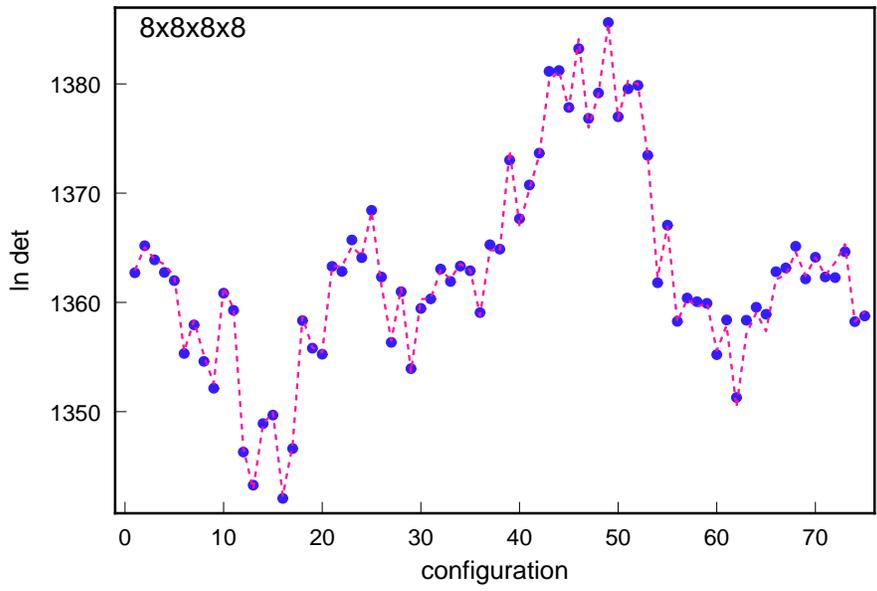,
width=0.8\hsize}
\caption{6 link fit  to ${\cal D}_{120}$, 8$^4$ lattices}
\label{fig:loops}
\end{figure}
\begin{figure}
\psfig{figure=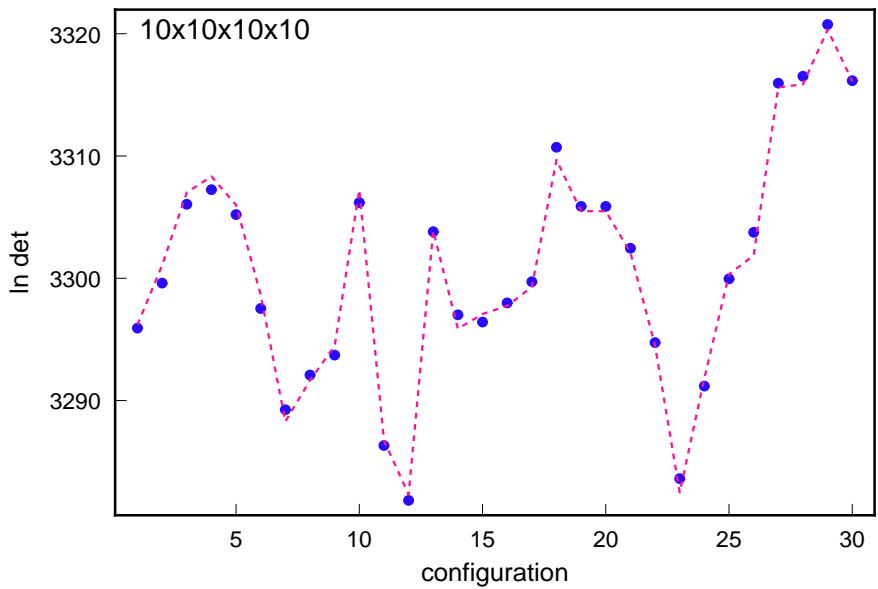,
width=0.8\hsize}
\caption{6 link fit  to ${\cal D}_{350}$, 10$^4$ lattices}
\label{fig:loops}
\end{figure}

  The accuracy of the loop fit to a truncated determinant is increased either by 
 including longer loops or by raising the IR cutoff (which requires that more  low
 eigenmodes be computed explicitly and included in the simulation). The
 variation with cutoff for the three different lattices studied is illustrated in Figure 5.
The eigenvalue cutoff has been reexpressed as a physical lattice momentum,
 with the conversion performed using an average spectral density obtained by
 averaging the individual spectra for all lattices in the ensembles used. Loops up to
 length 6, together with Polyakov lines stretching across the lattice, have been
 included in the fit of the determinant.
\begin{figure}
\psfig{figure=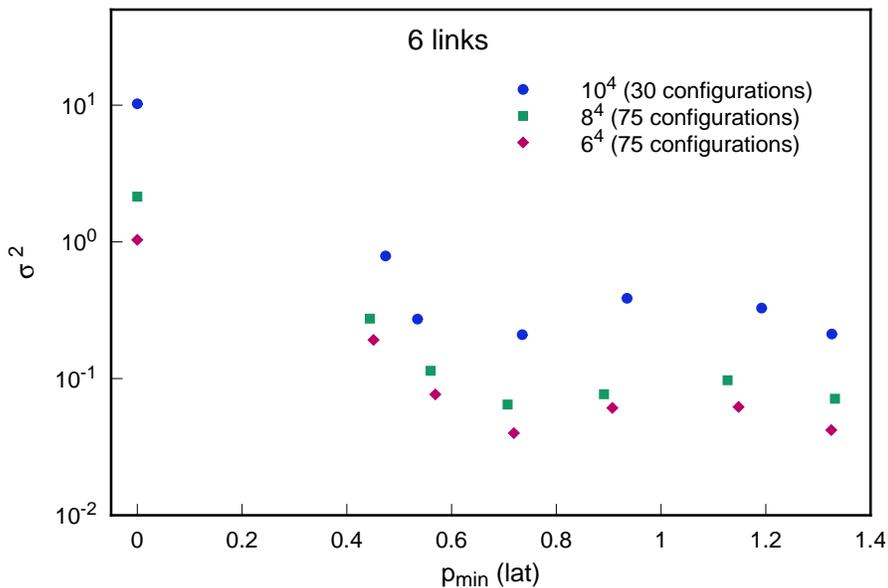,
width=0.8\hsize}
\caption{ $\sigma^{2}$ versus p$_{min}({\rm lat})$ for 6$^4$,8$^4$ and 10$^4$ lattices}
\label{fig:chisq6-8-10}
\end{figure}
The generic behavior is clear from Fig(5). The accuracy of the fit improves rapidly
with increasing cutoff  (initially, the variance decreases roughly like
 $e^{-Cp_{min}}$, where $p_{min}=|\lambda_{n_{\lambda}}|$ is the eigenvalue cutoff), reaching a small nonzero value.
For the lattices studied,the minimum attained $\sigma^{2}$  
 was less than one in all
 cases, leading to the close matching of actual and loop fit determinant values visible
 in Figures 2-4.  The $\sigma^{2}$ then fluctuates more or less randomly for further increases of cutoff around this value.  This small
 nonzero contribution at fixed $\beta$ (increasing with the lattice volume) is due to  higher dimension operators not included in
 the limited size loops in the fit and would presumably be reduced as the lattice $\beta$ is 
 increased to push the system towards the continuum. The fluctuations are simply a reflection
 of the limited statistics of our relatively small ensembles- they become more visible at
 larger cutoff where the signal to background is reduced. This can also be seen by
 plotting the cutoff dependence  of the $\sigma^{2}$ for the 6 link loop fits for the ensemble
 of 75 8$^4$ lattices broken into three subensembles of 25 configurations each (Fig 6).
 At smaller cutoffs, strong infrared correlations extending over several lattices lead to
 fairly large differences in the $\sigma^{2}$ of the fits for small p$_{\rm min}$, while at larger
 values of the cutoff, the size of fluctuations in each subensemble is comparable to
 the difference between the $\sigma^{2}$ for the different subensembles, suggesting that these
 fluctuations are statistical in origin.
 \begin{figure}
\psfig{figure=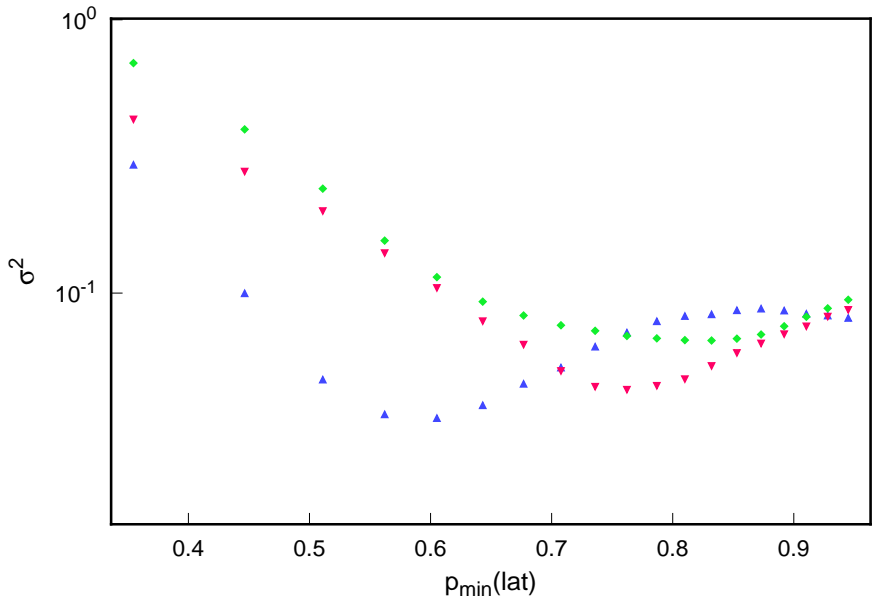,
width=0.8\hsize}
\caption{ $\sigma^{2}$ versus p$_{min}({\rm lat})$ for 3 ensembles of 25 8$^4$  lattices}
\label{fig:chisq_3set}
\end{figure}

  The accuracy of a simple loop representation for the ultraviolet part of 
the quark determinant, involving the
contribution of  a relatively small number of Wilson loop operators, depends on the
fact that only a few independent gauge-invariant operators exist of mass dimension 4
or 6 \cite{Gupta}. On the other hand,  the number of  topologically distinguishable Wilson loops grows much more rapidly (in fact, exponentially) with the length of the loops. This results in very
strong correlations between the values of distinct Wilson loop shapes over ensembles
of independent lattices. For example, the single plaquette value is strongly correlated
with the value of the 2x1 loop ($L_3$ in Figure 1). Thus, although the minimum 
$\sigma^{2}$ attainable for any given cutoff in link number is clearly only obtained by
employing all Wilson loops up to the prescribed length, the actual coefficients of
the individual loop values $L_{i}$ exhibit large variations from one subensemble to 
another  for a given lattice volume, and also as the lattice volume is increased. On the
other hand, if only truly independent operators are included, we should expect the
coefficents (properly normalized for the overall lattice volume) to remain relatively
constant  as the lattice volume is increased at fixed $\beta$, reflecting the contribution
of a definite set of low-dimension operators with expectation values approaching 
well-defined values in the infinite volume limit. 
\begin{table}
\begin{center}
\begin{tabular}{|c|c|c|c|}
\hline
Cutoff $(1/a)$ & Volume $N^4$ & \multicolumn{2}{c|}{coefficients}\\ 
$\Lambda_{QCD}$ & $N$ & constant & plaquette \\
\hline
 0.00 &  6  & -0.0225 & 0.0772 \\  
      &  8  & -0.0209 & 0.0794 \\  
\hline
 0.45 &  6  & -0.0111 & 0.0695 \\  
      &  8  & -0.0118 & 0.0699 \\  
\hline
 0.56 &  6  & -0.0043 & 0.0653 \\  
      &  8  & -0.0048 & 0.0653 \\  
\hline
\end{tabular}
\label{tbl:vol}
\caption{Variation with volume  of fit coefficients of link expansion of tr ln (Det) for various
low eigenvalue cutoffs $\Lambda$.}
\end{center}
\end{table}
This can be seen in Table 1, where we 
show the loop coefficients as a function of lattice volume for just the single plaquette operator obtained by minimizing the $\sigma^{2}$ with respect to a fit containing just a constant term
(the unit operator) and the single plaquette loops (in operator terms, $F_{\mu\nu}^{2}$). 
The 10$^4$ lattice results were not included because of the limited statistics available in this case 
(only 30 lattices as compared to 75 lattices for the 6$^4$ and 8$^4$ cases).

\section{ Quark determinant for heavy quarks-  hopping expansion vs effective loop actions}

   The main physical difference between the behavior of the quark determinants
 for light and heavy quarks lies in the relative variance of the infrared and
 ultraviolet contributions. For light quarks, there are important fluctuations 
 introduced both by the infrared modes, which incorporate the proper chiral
 behavior of the unquenched theory, and by the ultraviolet modes above
 $\Lambda_{QCD}$, which primarily renormalize the scale of the theory. Indeed,
 the latter are quantitatively dominant, but closely matched by an effective action
 involving only plaquettes or 6-link loops which for long distance physics
 reduces to an effective shift in the beta of the simulation. For heavy quarks,
 the density of the quark Dirac spectrum is much reduced in the infrared (which
 is cut off by the large bare quark mass), as are the fluctuations from the infrared modes
 below  $\Lambda_{QCD}$, while the UV modes still have substantial  variance, but
 again of a form which, as we shall show, can be accurately modeled by a simple loop action. 
 The relative variance of the low and high end contributions of the (log) quark determinant
 is shown in Fig 7, where the IR/UV cut is placed at the 15 th (positive or negative)
 mode, roughly at   $\Lambda_{QCD}$ for a 6$^4$ lattice at $\beta$=5.7 (for the
 sake of visibility, an irrelevant constant vertical offset has been applied to bring the
 various contributions close to zero). For the
 heavier quark, $\kappa=$0.1500, there is comparatively little variance in the
 IR contributions, but  large fluctuations in the UR part.

\begin{figure}
\psfig{figure=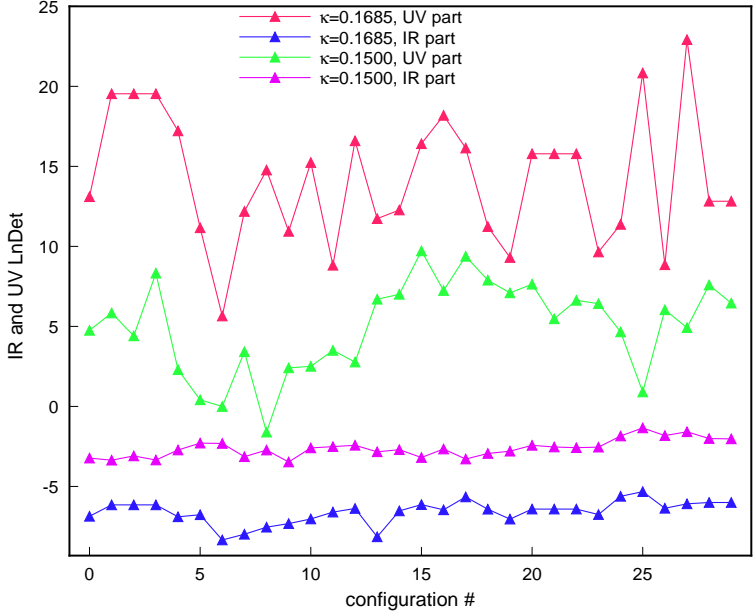,
width=0.7\hsize}
\caption{IR and UV contributions to quark determinants for light vs heavy quarks}
\label{fig:IR/UV}
\end{figure}

  The simplest approach one might take for the full quark determinant 
 for heavy quarks employs the well-studied \cite{Montvay}
 hopping parameter expansion for  $\cal{D}$=$\ln{(\gamma_{5}(D_W - m))}$, valid in the limit of
 small $\kappa$, i.e. large quark mass. In this section we examine the extent to which a truncated hopping
 parameter expansion can compete with the nonperturbative fitting of an effective
 loop action of the kind described in the preceding section. On a 6$^4$ lattice with
 periodic boundary conditions, we saw previously that there are five independent gauge invariant contributions
 to $\cal{D}$ of order $\kappa^6$ or less, denoted $L_{i},1\leq i \leq 5$ above. The loop averages $L_{i}(U)$ on a given gauge configuration $\{U\}$ are 
 normalized to give exactly one on the ordered configuration where all links are unity.
 Then a straightforward combinatoric exercise gives ($V$= lattice volume)
\begin{eqnarray}
 \cal{D} (U)&=& V(288\kappa^{4}L_{1}(U)-512\kappa^{6}L_{2}(U)+2304\kappa^{6}L_{3}(U)\\ \nonumber
  &+&4608\kappa^{6}L_{4}(U)+1536\kappa^{6}L_{5}(U)+O(\kappa^{8}))
\end{eqnarray}
 (Note that this computation gives an approximant to the {\em full} determinant, with no
 obvious way of implementing an IR/UV cutoff within the hopping parameter expansion.)

Unfortunately, the coefficients of higher order terms in the hopping parameter expansion 
 grow fairly rapidly (the number of closed loops increases exponentially with the 
 length of the loop) and the expansion converges slowly. For example, for $\kappa$=0.15,
 (with $\kappa^{8}$=2.6x10$^{-7}$ !) the average  $\cal{D}$ over an ensemble of 30
 configurations gives 172.4, while the hopping expansion gives 147.8. The discrepancy 
 is nevertheless mainly due to a few low dimension operators which dominate the contribution of
 the longer loops, as is apparent from Fig 8. The hopping parameter expansion through 6th
 order tracks roughly the exact determinants apart from an offset (the identity operator).
 Of course, the nonperturbative fit including $L_{1-5}(U)$ does much better 
(the $\sigma^{2}$ is 0.112).
\begin{figure}
\psfig{figure=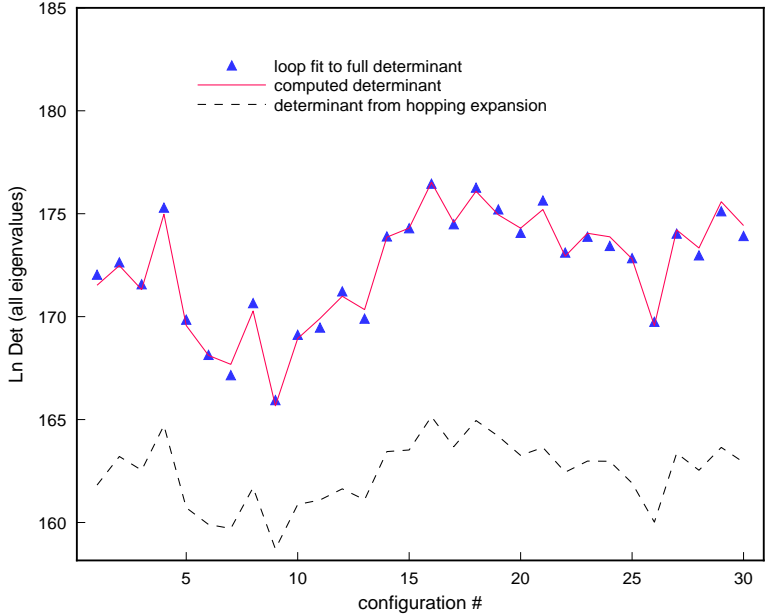,
width=0.7\hsize}
\caption{Full determinant, hopping parameter vs nonperturbative fit}
\label{fig:detfit}
\end{figure}
We can improve the agreement of the truncated hopping expansion with the data
 by varying both the offset and the single plaquette component  from the value of the $\kappa^4$
 coefficient given in Eq (2). This amounts to including at least the lowest nontrivial
 dimension operator (dimension 4) arising from the longer loops (length 8 and
 higher).  This fit is shown in Fig 9, in comparison with the unmodified hopping
 expansion result shifted only by a constant offset. The fit is certainly improved by 
 optimizing the single plaquette component (the $\sigma^{2}$ is now  0.75) but clearly
 the nonperturbative fit of Fig. 8, in which all closed loops through length 6 are
 optimized,  still wins by a substantial factor. We can conclude that even for quite
 heavy quarks, the hopping parameter expansion, though analytically available, is
 not competitive with a nonperturbative fit using even a small number of Wilson loops.
 Determining the coefficients in such a fit only requires the extraction
 of the determinant for a few typical configurations. We find that a sufficiently
 accurate determination of the loop coefficients can be obtained by calculating
 the spectrum on as few as 10 gauge configurations.
\begin{figure}
\psfig{figure=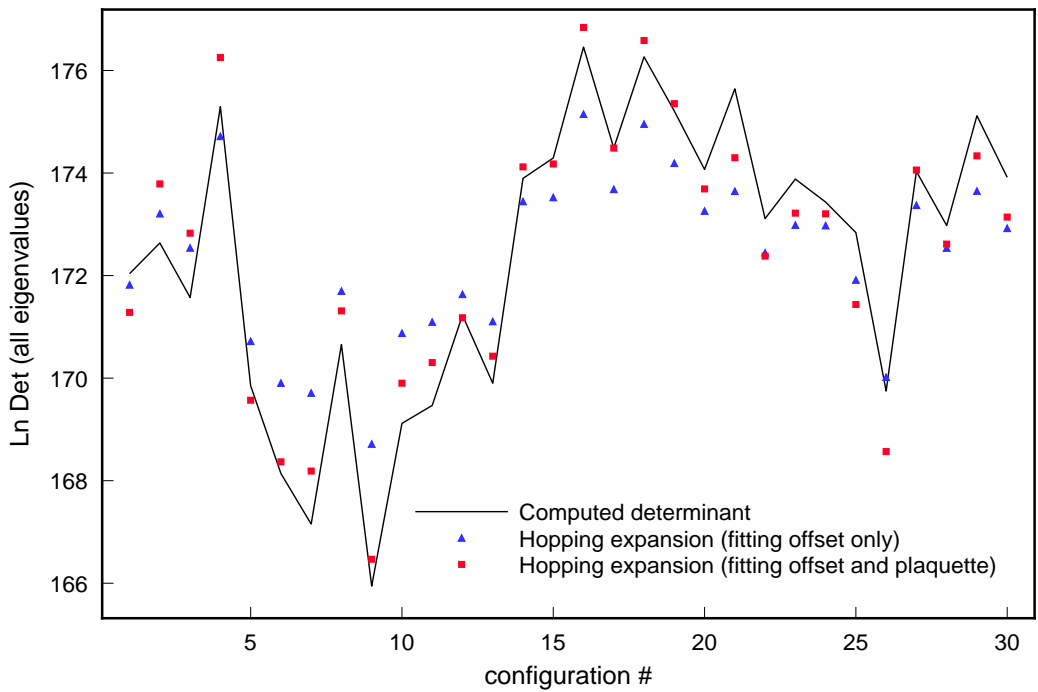,
width=0.95\hsize}
\caption{Hopping expansion fits, with offset vs with offset+plaquette}
\label{fig:plaqonly}
\end{figure}

   We pointed out earlier that the IR portion of the Dirac spectrum is relatively inert in
 the case of heavy quarks. One therefore expects that a nonperturbative fit with a
 few small loops should be accurate for heavy quarks, even if one insists on fitting the full determinant,
 as discussed above. Indeed, we saw above that the $\sigma{2}$ with a fit to the
 full determinant is 0.112 including loops up to length 6. The fit is still improved by
 excising the IR part of the spectrum, though not as dramatically as in the case of light
 quarks where the IR fluctuations require the inclusion of large loops. For $\kappa$=0.15,
 one finds for example that the $\sigma^{2}$ of a 5 loop fit decreases to 0.035,  0.026 and
 0.019 when the lowest 30, 60 or 90 modes are excluded from the determinant.

\section{Extracting the complete Dirac spectrum- explicit versus stochastic approaches}

  A fairly efficient method for performing a complete spectral resolution of the hermitian 
 Wilson-Dirac operator was described some time ago by Kalkreuther \cite{lanczos}. One
 employs the usual Lanczos procedure, {\em without} reorthogonalization, pruning out
 the spurious eigenvalues by the Cullum-Willoughby procedure \cite{Cullum}. Typically
 the extraction of the complete Dirac spectrum requires (due to the inexact arithmetic)
 {\em more} Lanczos sweeps than the actual dimension (12$V$, where $V$ is the lattice
 volume), by a factor of 2-3. For example, on a 10$^4$ lattice, the full spectrum is
 obtained after about 320000 Lanczos sweeps, as compared to the actual dimension
 of 120000.  The convergence of the procedure is improved by starting from gauge-fixed
 configurations. (In spite of the fact that the spectrum is gauge-invariant, it appears that the
 presence of gauge noise can reduce the numerical stability of the Lanczos procedure.)
 Occasionally, spectral fluctuations lead to two eigenvalues which are almost degenerate,
 or a real eigenvalue almost degenerate with a spurious one,
 and the spectrum is found to be missing a small number of eigenvalues (note that the
 Lanczos procedure does not identify the degeneracy of the various eigenvalues). In an
 ensemble of 75 6$^4$ lattices, the procedure missed a single mode in only two cases.
 For an ensemble of 20 10$^4$ lattices, 3 eigenvalues were missed 5 times, 2 for 5
 configurations, 1 for 6 configurations, and complete spectra were obtained for 4
 configurations.  Although the procedure can probably be tuned to reduce the
 frequency of missed eigenvalues, the general trend is nevertheless towards troublesome
 accidental degeneracies for larger lattices. Recently, we have found that the problem
 on larger lattices can be considerably ameliorated by a more careful retuning of the
 algorithm for identifying spurious eigenvalues- the resulting complete spectra agree
 completely with the reconstructed eigenvalues discussed below. Moreover, the convergence
 of the eigenvalues in the densest part of the spectrum occurs very rapidly as one reaches the
 end of the procedure, so that the spectrum of the tridiagonal matrix essentially consists of
 either spurious eigenvalues (easily identified by the Cullum-Willoughby procedure) or
 accurate converged eigenvalues. For example, with 320000 Lanczos sweeps on a 10$^4$
 lattice (120000 eigenvalues) one finds with a carefully tuned Lanczos calculation exactly
 200000 spurious eigenvalues and 120000 converged eigenvalues, with the latter satisfying the
 set of exact sumrules discussed below to high accuracy (for a 10$^3$x20 lattice
 with 240000 eigenvalues, complete spectra are obtained with 600000 Lanczos sweeps).
 The present tuning leaves about 3
 orders of magnitude between the tolerance test for spurious eigenvalues (set at 10$^{-11}$)
 and the arithmetic precision (about 10$^{-14,15}$).  For larger lattices where the volume and hence
 the maximum spectral
  density is a few orders of magnitude higher, we expect the Cullum-Willoughby procedure to misidentify
 some true eigenvalues as spurious, leading to incomplete spectra (as we indeed find if the tolerance
 parameter for spurious eigenvalues is set to 10$^{-10}$, for example).

   In fact, in all cases mentioned above, incomplete spectra can be completely repaired with
 the aid of  exact sum rules for traces of powers of the Wilson-Dirac matrix $H=\gamma_{5}(D_W - m)$. 
 As we lose at most 3 eigenvalues, the lowest four sum rules suffice to determine all missing
 eigenvalues, with a spare relation left over for checking the accuracy of the
 reconstruction.  One easily derives 
\begin{eqnarray}
   {\rm Tr}(H) &=& 0 \\
   {\rm Tr}(H^2) &=& 12V(1+16\kappa^{2}) \\
   {\rm Tr}(H^3) &=& 0 \\
   {\rm Tr}(H^4) &=&  12V(1+64\kappa^{2}+(448-96<P>)\kappa^{4})
\end{eqnarray}
 where the Wilson-Dirac operator $D_W-m$ in $H$ is normalized to be of the form
 $1 +\kappa({\rm link\;\;terms})$, and $<P>$ in Eq(7) is the average plaquette value in
 the given configuration. A check of these sum rules on  configurations where the
 entire spectrum has been successfully extracted gives ${\rm Tr}(H,H^3)<$10$^{-7}$ while the quadratic and quartic sum rules are reproduced to at least eight significant figures.
  Repairing the incomplete spectra with these sum rules reveals,
 as expected, that the missing eigenvalues are either close to degenerate with each other
 or with a converged eigenvalue from the Lanczos procedure. 

 An alternative to the explicit spectral resolution of  $H$, which is only really feasible for
 small to moderate sized lattices, is the stochastic approach developed by Golub
 and coworkers \cite{golub1,golub2}, and applied by Irving and Sexton in their study \cite{IrvSext} of
 loop actions for the quark determinant. Their formalism uses the close connection between the
 Lanczos recursion and Gaussian integration to generate rigorous lower and upper
 bounds to the diagonal matrix element  $<v|f(A)|v>$ of any differentiable function $f(A)$
 of a positive definite matrix $A$.  The spectral sum giving this matrix element is
 transformed to a Riemann-Stieltjes integral and the usual quadrature rules (Gauss,
 Gauss-Radau, Gauss-Lobatto, etc.) applied to this integral can then be reexpressed
 in terms of a Lanczos recursion (for further details, we refer the reader to the paper
 of  Bai, Fahey and Golub \cite{golub2}).  The use of alternative quadrature rules
 (in which information about the upper and lower limits of the spectrum is included
 in the Gaussian measure) does not seem to matter much in the application to the
 Wilson-Dirac matrix, although we have found that the Gauss-Lobatto version 
 requires about  50\% more Lanczos sweeps to achieve the same precision
 as the other quadrature rules.  It is straightforward to generalize the 
 arguments of \cite{golub2} to show that the Lanczos estimates converge to the correct
 answer even for non-positive definite hermitian matrices (such as the hermitian 
 Wilson-Dirac operator $H$), although the strict upper and lower bounds provided
 by the formalism no longer hold in the case $f(A)=\ln(A)$, as they depend on 
 positivity of the derivatives of $f(A)$ over the entire spectrum. The rate of  convergence of this procedure 
 is impressive- for an 8$^4$ lattice the estimate of  $<v|\ln{|H|}|v>$ for a generic random
 vector $v$ with only  300 Lanczos sweeps is accurate to 7 places on a 8$^4$ lattice, and
 to about 5 places on a 12$^3$x24 lattice.

  The Gaussian/Lanczos approach described above immediately leads to a stochastic
 method for estimating $\ln {\rm det}(|H|)=\sum_{i}<v_{i}|\ln{|H|}|v_{i}>$, where the sum 
 extends over a complete orthonormal basis. Namely, one computes an average
 over a set of random vectors $|z_i>$ where each vector has components $\pm 1$
 chosen at random:
\begin{eqnarray} 
  \ln {\rm det}(|H|)&=& E(<z|\ln{|H|}|z>)  \\
  var(<z|\ln{|H|}|z>) &=& 2\sum_{i \not= j} |\ln{|H|}_{ij}|^{2}
\end{eqnarray}
 We have checked that the choice of elements $\pm 1$ for the components of the
 random vectors is optimal, in the sense that the variance (8) obtained with this
 choice cannot be further reduced by an alternative choice of random variable.
 This procedure therefore gives an unbiased estimator for the full quark determinant,
 with errors that are purely statistical, decreasing as $1/\sqrt{N}$ with the number $N$ of
 random vectors used. Evidently the accuracy achieved for a given amount of
 computational effort is directly determined by the size of the offdiagonal matrix
 elements of $\ln{|H|}$. This is of course a very complicated functional of the
 gauge field for a general configuration. For the ordered configuration, however,
 $H$ can be explicitly diagonalized, and a straightforward computation yields 
\begin{eqnarray}
  var(<z|\ln{|H|}|z>) &=& 4V\{\frac{1}{V}\sum_{k}\ln{(A(k)^{2}+B_{\mu}(k)B_{\mu}(k))}^{2} \nonumber \\
 &-&(\frac{1}{V}\sum_{k}\ln{(A(k)^{2}+B_{\mu}(k)B_{\mu}(k))})^{2} \}
\end{eqnarray}
where $A(k)=1-2\kappa\sum_{\mu}\cos{(k_{\mu})}, B_{\mu}(k)=2\kappa\sin{(k_{\mu})}$,
 and $V$ is the lattice volume. In other words, the variance of the estimates (8)
 is exactly  equal to the variance in the logarithm of the free quark lattice offshellness
 (as $A(k)^{2}+B_{\mu}(k)B_{\mu}(k)$ is just the lattice version of $k_{\mu}k_{\mu}+m^2$
 in the continuum).  For not too heavy quarks (it is easy to see that
 the variance (9) vanishes with $\kappa$)
 the number multiplying the lattice volume $V$ in (9) is of order unity, so the optimal
 stochastic procedure requires $N \simeq \sqrt{V}$ to determine $ \ln {\rm det}(|H|)$ to within an
 additive error of order unity, assuming the free quark case as a rough guide. 
 Explicit calculations show that the order of magnitude of the variance is indeed as given
 in (9). For example, the free quark 
 formula (9) gives about 8x10$^3$ for the variance on an 8$^4$
 lattice (choosing $\kappa$=0.12), while an actual run with 1000 random vectors
 on a nontrivial 8$^4$ configuration ($\beta$=5.7, $\kappa$=0.1685) gives a
 variance of 9.7x10$^3$, and a final result for the mean (log) determinant is  1224.0$\pm$ 3.1 (the error is obtained by taking the square root of the variance per data
 point).
 This should be compared with the exact value obtained from the complete spectral
 resolution carried out by the direct Lanczos approach described in the first
 part of this section, which was 1222.148.
 On a 12$^3$x24 lattice at $\beta$=5.9, $\kappa=$0.1597, a typical configuration
 gave a variance of 8.0x10$^4$, compared with 7.2x10$^4$ from the free quark
 result (9) (again using $\kappa$=0.12 for the free case). An explicit evaluation with 820
 random vectors gave a final result 8682$\pm$ 9.9 for the (log) determinant in this case.
 
   We are now in a position to compare the computational efficiency of the direct and 
 stochastic methods described above.  The stochastic approach yields an estimate
 of the logarithmic determinant  accurate to any fixed preassigned error  with a 
 computational effort growing like $V^{3/2}$, while the full Lanczos spectral resolution,
 which effectively determines the determinant to machine precision (actually, about
 eight significant figures), requires an effort
 of order $V^2$.   However, the prefactors in each case render the Lanczos approach
 advantageous for small to moderate (say, 12$^4$) lattices. For example, on an
 8$^4$ lattice, the complete spectral resolution requires on the order of 130000 applications
 of the ``dslash" (lattice covariant quark derivative) operator,  while the stochastic
 method would require about 10000 random vectors, each with 200 Lanczos sweeps-
 i.e a total of  2 million dslash operations- to reduce the error on the determinant to order 
 unity (four significant figures). On the other hand, as discussed previously, on large lattices the direct Lanczos procedure  is likely to fail as  the machine precision will be inadequate to resolve the increasing number
 of accidental degeneracies due to the high spectral density.  In this case the stochastic 
 approach may be the only option for estimating the complete quark determinant.

\vspace{1in}
\section{Acknowledgements}
 The work of A.D. was supported in part by NSF grant 97-22097.
 The work of  E.E was performed at the Fermi National Accelerator Laboratory,
 which is operated by Universities Research Association, Inc., under contract
 DE-AC02-76CHO3000.  
 The work of H.T. was supported in part by the Department of Energy under 
 grant DE-AS05-89ER 40518.

 \end{document}